\documentstyle[11pt,epsf]{article}

\catcode`\@=11 
\def\lsim{\mathrel{\mathpalette\@versim<}}
\def\gsim{\mathrel{\mathpalette\@versim>}}
\def\@versim#1#2{\vcenter{\offinterlineskip
        \ialign{$\m@th#1\hfil##\hfil$\crcr#2\crcr\sim\crcr } }}
\catcode`\@=12 

\begin{document}
\begin{titlepage}
  \begin{flushright}
    KUNS-1534 \\[-1mm]
    HE(TH)98/15 \\[-1mm]
    hep-ph/9810204
  \end{flushright}
  \begin{center}
    \vspace*{1cm}

    {\Large\bf Right-handed neutrino mass and bottom-tau ratio\\ in
      strong coupling unification}
    \vspace{1cm}

    {Masako {\sc Bando}\footnote{E-mail address:
        bando@aichi-u.ac.jp} and
      Koichi {\sc Yoshioka}\footnote{E-mail address:
        yoshioka@gauge.scphys.kyoto-u.ac.jp}}
    \vspace{5mm}

    $^*${\it Aichi University, Aichi 470-02, Japan} \\
    $^{\dagger}${\it Department of Physics, Kyoto University
      Kyoto 606-8502, Japan}
    \vspace{1cm}

    \begin{abstract}
      The recent results of neutrino experiments indicate the
      existence of the right-handed neutrinos with masses around
      intermediate scale, which affects the prediction of the
      bottom-tau mass ratio. In the minimal supersymmetric standard
      model, this effect largely depends on the right-handed neutrino
      mass scale and then may severely limit its lower bound. In this
      letter, we show that in the case of strong coupling unification
      the bottom-tau mass ratio is little affected by the presence of
      the right-handed neutrinos. This is because of the infrared
      fixed-point behavior of Yukawa couplings, which is a common
      feature of this kind of model.
    \end{abstract}
  \end{center}
\end{titlepage}
\newpage

The gauge coupling unification is one of the remarkable successes of
the minimal supersymmetric standard model (MSSM) \cite{unify} and
provides us with a strong phenomenological support to the MSSM\@. The
bottom to tau mass ratio, $R$ $(\equiv y_b/y_\tau)$, is another
phenomenological support of the MSSM and has been intensively studied
\cite{b-tau}\@. With the $SU(5)$ unification condition of Yukawa
couplings, $y_b=y_\tau$, the $SU(3)_C$ gauge coupling effect gives
$R(M_Z)\sim 2.2$, which was almost a desirable low-energy
value. However, a more accurate experimental data on the bottom and
tau masses \cite{exp} gives
\begin{eqnarray}
  R(M_Z) \,\simeq\, 1.6 - 1.8,
  \label{ratio}
\end{eqnarray}
which strongly constrains the allowed region of relevant parameter
space. In order to reproduce the above ratio, it is known that the top
(and/or bottom) Yukawa coupling should be very large so that it
suppresses the $SU(3)_C$ gauge coupling effect \cite{b-tau}.

Now, the recent neutrino experiments strongly indicate the existence
of right-handed neutrinos. In particular, the atmospheric neutrino
anomaly \cite{atm} can be solved by neutrino oscillation scenario
between two neutrino species with squared mass difference $\sim
10^{-(2-3)}$ eV$^2$\@. By the seesaw mechanism \cite{seesaw}, this
fact suggests that the right-handed neutrino mass scale $M_R$ is of an
intermediate scale. The existence of right-handed neutrinos at an
intermediate scale is also favored by cosmology and astrophysics from
various points of view (hot dark matter \cite{HDM}, baryogenesis
\cite{baryon}, inflation \cite{inflation}, etc.)\@. However, in the
MSSM, if the third generation right-handed neutrino exists at such an
intermediate scale we have a following 1-loop renormalization group
equation (RGE) for $R$ in the region between the GUT scale $M_G$ and
$M_R$:
\begin{eqnarray}
  \frac{dR}{dt} \,=\, \frac{R}{16\pi^2} \left[
    (y_t^2-y_\nu^2) +3(y_b^2-t_\tau^2) -\left(\frac{16}{3}g_3^2
      -\frac{4}{3}g_1^2\right) \right],
 \label{R}
\end{eqnarray}
where $g_i\, (i=1,2,3)$ are gauge couplings of the standard gauge
group $SU(3)_C\times SU(2)_W\times U(1)_Y$, and $t \equiv \ln
\mu$\@. In the above equation, a new ingredient from $y_\nu$, which is
absent in the usual MSSM, gives an opposite contribution to that of
the top Yukawa coupling. If one assumes the $SO(10)$-like boundary
condition, $y_t=y_\nu$, the neutrino Yukawa coupling largely reduces
the effect of the top Yukawa coupling and increases the low-energy
value of the bottom-tau ratio. This effect of $y_\nu$ imposes the
strong restriction on the allowed region of $M_R$ and/or $\tan \beta$
\cite{smirnov,murayama}. In particular, in the small $\tan \beta$
case, the severe constraint for the lower bound of $M_R$ is found. If
one considers the tau-neutrino as a hot dark matter candidate the
lower bound of $M_R$ translates into an upper bound on the neutrino
hot dark matter density of the universe, which is much smaller than
the cosmologically interesting range \cite{murayama}.

So far, there are several ways to avoid this situation. Within
the MSSM framework, one way is to modify the right-handed neutrino
Majorana mass matrix, in which the mass of the third-generation
right-handed neutrino is much larger than the intermediate scale. This
can be achieved consistently with the quark-lepton parallelism
preserving the large mixing between the second and third generation
neutrinos \cite{Majorana}. The GUT scale physics may also change the
bottom-tau ratio by modifying the boundary condition, $y_b=y_\tau$,
with relevant Higgs fields \cite{boundary}, by the corrections from
GUT scale physics \cite{GUTb-tau,murayama}, and so on. If one can
adopt the models with an intermediate gauge group beyond the standard
model \cite{inter}, the boundary condition, $y_b=y_\tau$, preserves
down to a breaking scale of an intermediate gauge group and the
right-handed neutrinos, which get masses of this breaking scale, give
no harmful effect on the low-energy bottom-tau ratio.

In this letter, we suggest an alternative approach to this problem in
the framework of the standard gauge symmetry. We point out that in
strong coupling unification scenario such a bound on $M_R$ does not
exist in contrast to the MSSM case (weak coupling unification)\@. In
the strong coupling unification scenario \cite{strong}, the gauge
couplings behave asymptotically non-freely yielding a strong unified
gauge coupling ($\sim O(1)$)\@. For a concrete model, we here take the
extended supersymmetric standard model (ESSM) with 5 generations; the
MSSM + 1 extra vector-like family \cite{essm,BOST}, as a typical
example of the strong coupling unification model. However, the results
are not specific to the ESSM but common features in general strong
unification models.

In the ESSM, the RGEs of the Yukawa couplings at 1-loop level are as
follows:
\begin{eqnarray}
  \frac{dy_t}{dt} &=& \frac{y_t}{16\pi^2} \left(9y_t^2+y_b^2+2y_\nu^2
    -\frac{16}{3}g_3^2 -3g_2^2 -\frac{13}{15}g_1^2 \right),
  \label{top} \\
  \frac{dy_b}{dt} &=& \frac{y_b}{16\pi^2} \left(y_t^2+9y_b^2+2y_\tau^2
    -\frac{16}{3}g_3^2 -3g_2^2 -\frac{7}{15}g_1^2 \right), \\
  \frac{dy_\tau}{dt} &=& \frac{y_\tau}{16\pi^2} \left(6y_b^2+5y_\tau^2
    +y_\nu^2 -3g_2^2 -\frac{9}{5}g_1^2 \right), \\
  \frac{dy_\nu}{dt} &=& \frac{y_\nu}{16\pi^2} \left(6y_t^2 +y_\tau^2
    +5y_\nu^2 -3g_2^2 -\frac{3}{5}g_1^2 \right).
  \label{neu}
\end{eqnarray}
Here, we consider only the 3rd and 4th generation Yukawa couplings and
neglect the other generations (throughout this letter, we set the mass
of the extra families to be the same order of the SUSY breaking scale
$M_S$ ($\simeq 1$ TeV) \cite{essm,BOST})\@. This assumption correctly
reproduces the low-energy top quark mass as its infrared fixed-point
value. Up to this level there is no difference between the MSSM and
the ESSM in the expression of the RGE for $R$ (eq.\ (\ref{R})). Note
that the other generation Yukawa couplings, if included, affects the
RGE for $R$ very little since they are very small or come in below
2-loop level.

There are two characteristic features in the ESSM which are important
in investigating the behavior of the low-energy bottom-tau ratio. One
is that, due to the strong gauge couplings, some of the Yukawa
couplings (over gauge coupling) at low-energy scale are determined
almost insensitively to their initial values at $M_G$ because they
reach their infrared fixed points very rapidly \cite{IRFP}. This is
physically significant because we can get important information on the
physical parameters independently of unknown high-energy physics. The
other is the boundary condition of the bottom and tau Yukawa
couplings. Generally, one must adopt the unification condition $y_\tau
> y_b$ in such a strong unification model because of the strong
enhancement effect on the bottom-tau ratio from $SU(3)_C$ gauge
interaction. In the ESSM, we impose the boundary condition of the
Yukawa couplings as $y_\tau=3y_b$ at $M_G$ by assuming the Higgs field
of the $\overline{126}$ ($\overline{45}$) representation of $SO(10)$
($SU(5)$)\@. With this boundary condition, we showed that the correct
observed bottom-tau mass ratio can be reproduced when we do not
include the effects of neutrino Yukawa couplings \cite{BOST}.

To see these two points more concretely, it is instructive to use the
semi-analytic expression of the low-energy value of $R$\@. If all the
Yukawa couplings are neglected, by formally integrating the 1-loop RGE
(eq.\ (\ref{R})), we have
\begin{eqnarray}
  \frac{R(M_Z)}{R(M_G)} &=& f_R \left(
    \frac{\alpha_3(M_S)}{\alpha_3(M_G)} \right)^{8/9} \left(
    \frac{\alpha_1(M_S)}{\alpha_1(M_G)} \right)^{10/99}, \qquad
  {\rm (MSSM)} \label{enhance-M} \\
  \frac{R(M_Z)}{R(M_G)} &=& f_R \left(
    \frac{\alpha_3(M_S)}{\alpha_3(M_G)} \right)^{-8/3} \left(
    \frac{\alpha_1(M_S)}{\alpha_1(M_G)} \right)^{10/159}, \qquad
  {\rm (ESSM)} \label{enhance-E}
\end{eqnarray}
where $f_R$ is an enhancement factor of $R$ which comes from the net
contributions between $M_S$ and $M_Z$ including the effect of the top
Yukawa and $SU(3)_C$ gauge couplings ($f_R \sim 1.2$)\@. In the MSSM
case (\ref{enhance-M}), we have $R(M_Z)/R(M_G) \sim 2.4$ for
$\alpha_3(M_Z)=0.12$, which is somewhat larger than the present
experimental value. On the other hand, in the ESSM case
(\ref{enhance-E}), the strong unified gauge coupling considerably
enhances $R(M_Z)$\@. However, as mentioned above, Yukawa couplings
evolve up to their theoretical infrared fixed points very quickly just
below the GUT scale in the ESSM\@. Including the effects of Yukawa
couplings as their infrared fixed-point values which are obtained from
eq.\ (\ref{top}), we have
\begin{eqnarray}
  \frac{R(M_Z)}{R(M_G)} &\simeq& f_R \left(
    \frac{\alpha_3(M_S)}{\alpha_3(M_G)} \right)^{-7/5} \left(
    \frac{\alpha_1(M_S)}{\alpha_1(M_G)} \right)^{10/159}. \qquad
  {\rm (ESSM)} \label{enhance-E2}
\end{eqnarray}
This factor gives $R(M_Z)/R(M_G) \sim (5-6)$ (for
$\alpha_3(M_Z)=0.12$)\@. Therefore, we can get a correct low-energy
bottom-tau ratio if we take the boundary condition, $y_\tau=3y_b$
$(R(M_G)=1/3)$, which can be naturally obtained from a single relevant
Higgs field ($\overline{126}$ ($\overline{45}$) of $SO(10)$ ($SU(5)$))
\cite{boundary}. Then, it is our task to see how the right-handed
neutrinos affect on this successful prediction of the bottom-tau ratio
in strong unification models.

\begin{figure}[htbp]
  \begin{center}
    \leavevmode
    \epsfxsize=9cm \ \epsfbox{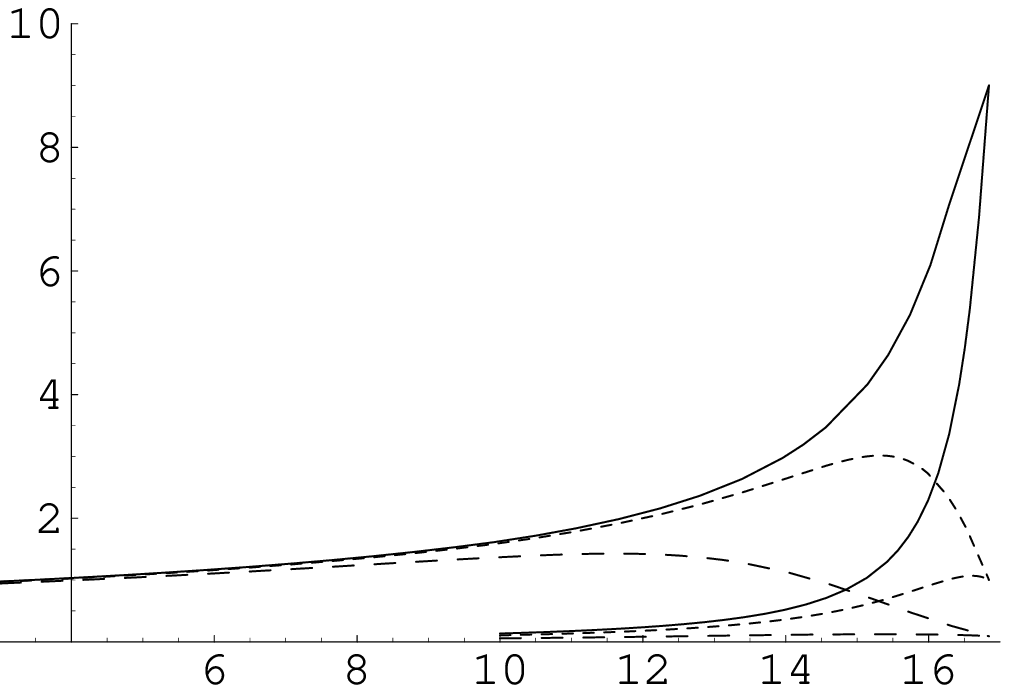}
    \put(-242,188){$y_{t,\nu}^2$}
    \put(-33,115){$y_t^2$}
    \put(-2,100){$y_\nu^2$}
    \put(7,13){$\ln_{10}\mu$}
    \caption{The behavior of the running couplings
      of $y_t$ and $y_\nu$ in the ESSM for the initial values,
      $y_t=y_\nu=0.3,1$ and 3 at $M_G$.}
    \label{fig:tnu-essm}
    \vspace{10mm}
    \epsfxsize=9cm \ \epsfbox{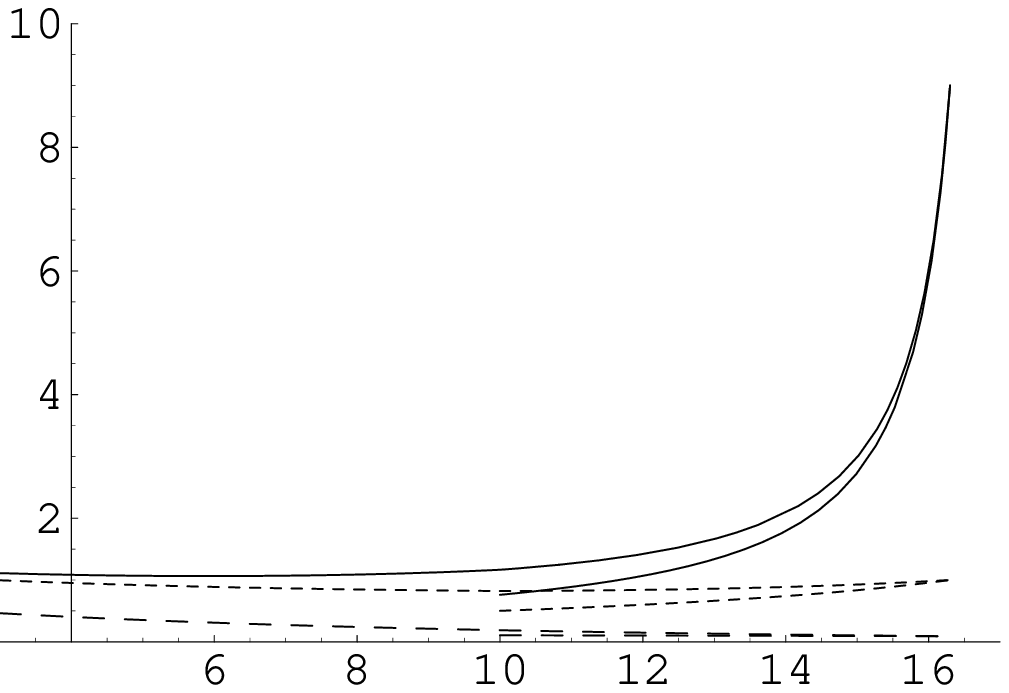}
    \put(-242,188){$y_{t,\nu}^2$}
    \put(-33,120){$y_t^2$}
    \put(-10,110){$y_\nu^2$}
    \put(7,13){$\ln_{10}\mu$}
    \caption{The behavior of the running couplings
      of $y_t$ and $y_\nu$ in the MSSM for the initial values,
      $y_t=y_\nu=0.3,1$ and 3 at $M_G$. }
    \label{fig:tnu-mssm}
  \end{center}
\end{figure}

With the common expression of the RGE for $R$, however, there are two
notable differences in the cancellation effect between Yukawa
couplings because of the above mentioned two characteristic features
of the ESSM; the infrared fixed-point structure of (top) Yukawa
coupling and the boundary condition of the bottom and tau Yukawa
couplings. Especially, due to the latter effect, the neutrino Yukawa
coupling $y_\nu$ decreases more rapidly as $\mu$ decreases from
$M_G$\@. This is because the effect of Yukawa couplings is very large
because of the boundary condition $y_\tau = 3y_b$ together with their
large coefficients in the beta function (\ref{neu})\@. These two
effects on the cancellation between $y_t$ and $y_\nu$ can be seen from
Fig.\ \ref{fig:tnu-essm}\@. For small $y_G$ (the initial value of the
Yukawa couplings at $M_G$ ($y_t=y_\nu\equiv y_G$)), $y_t$ grows up
quickly due to the former effect while in the case of large $y_G$,
$y_\nu$ decreases quickly due to the latter effect. In any case, there
is a striking difference between $y_t$ and $y_\nu$ just below the GUT
scale. Thus, the cancellation effect between $y_t$ and $y_\nu$ in the
first bracket in eq.\ (\ref{R}) becomes much weaker and we can expect
that the existence of the right-handed neutrinos does not affect the
low-energy bottom-tau ratio so seriously. On the other hand, in the
MSSM case, $y_t$ and $y_\nu$ behave almost similarly and the
cancellation causes important effects on the bottom-tau ratio, as is
well known (Fig.\ \ref{fig:tnu-mssm})\@.

\begin{figure}[htbp]
  \begin{center}
    \leavevmode
    \epsfxsize=10cm \ \epsfbox{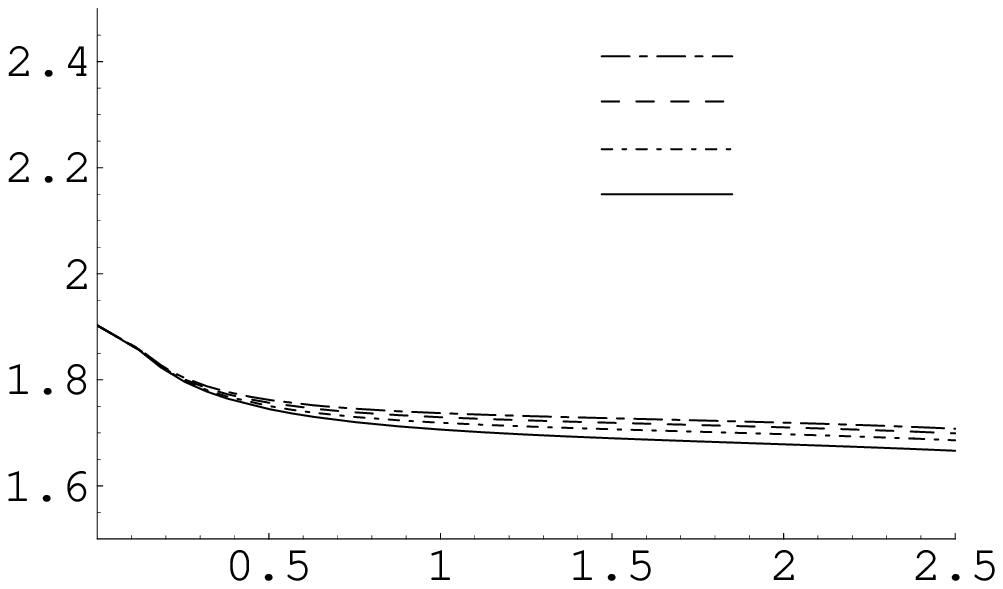}
    \put(-270,185){$R(M_Z)$}
    \put(5,25){$y_t(M_G)$}
    \put(-67,160){\small $M_R=10^{10}$ GeV}
    \put(-67,147){\small $M_R=10^{12}$ GeV}
    \put(-67,134){\small $M_R=10^{14}$ GeV}
    \put(-67,121){\small $M_R=M_G$}
    \caption{An example of the $M_R$ dependence of the predicted value
      $R(M_Z)$ in the ESSM (for $\alpha_3(M_Z)=0.12$).}
    \label{fig:r-essm}
    \vspace{10mm}
    \epsfxsize=10cm \ \epsfbox{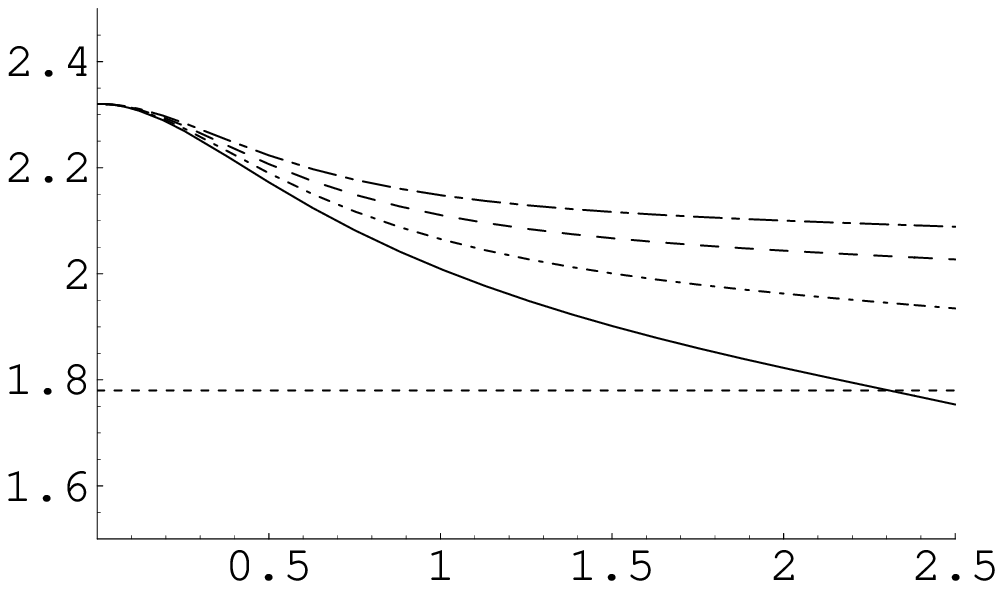}
    \put(-270,185){$R(M_Z)$}
    \put(5,25){$y_t(M_G)$}
    \caption{An example of the predicted value of $R(M_Z)$ in the MSSM
      (for $\alpha_3(M_Z)=0.12$)\@. The values of $M_R$ are as in
      Fig.\ 3\@. The dotted line shows the experimental upper bound on
      $R(M_Z)$.}
    \label{fig:r-mssm}
  \end{center}
\end{figure}

Next, let us numerically calculate the bottom-tau mass ratio at
low-energy scale with those running Yukawa coupling constants at
hand. As we have already noted, in the ESSM the top and bottom Yukawa
couplings reach very rapidly their infrared fixed points whose value
are almost equal each other irrespectively to their initial boundary
conditions. Thus we are inevitably lead to a large $\tan \beta$ case
except for very hierarchical bottom Yukawa coupling. To see the
differences between two models more clearly, it is useful to show the
obtained result of $R(M_Z)$ versus the initial condition of $y_t$ for
various values of $M_R$ (Figs.\ \ref{fig:r-essm} and
\ref{fig:r-mssm})\@. In these figures, we have used the 2-loop RGEs
and included the 1-loop threshold corrections at the SUSY breaking
scale, which may be important for the large $\tan \beta$ case
\cite{threshold}. In the MSSM case (Fig.\ \ref{fig:r-mssm})
\cite{smirnov,murayama}\@, it is clear that there are appreciable
$M_R$ dependence of $R(M_Z)$ and the bounds on $M_R$ do
exist. Compared with this result, we clearly see a much weaker
dependence of $M_R$ in the ESSM case (Fig.\ \ref{fig:r-essm}) as we
have expected from the running Yukawa couplings of Fig.\
\ref{fig:tnu-essm}\@.

\begin{figure}[htbp]
  \begin{center}
    \leavevmode
    \epsfxsize=10cm \ \epsfbox{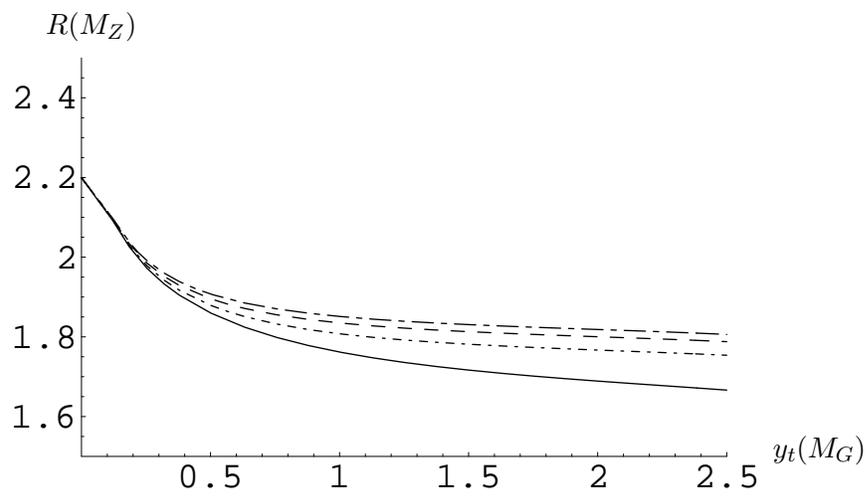}
    \put(-270,185){$R(M_Z)$}
    \put(5,25){$y_t(M_G)$}
    \caption{An example of the predicted value of $R(M_Z)$ in the ESSM
      for small $\tan \beta$ case (for $\alpha_3(M_Z)=0.12$)\@. The
      values of $M_R$ are as in fig.\ 3.}
    \label{fig:r-essm2}
  \end{center}
\end{figure}

{}From Figs.\ \ref{fig:r-essm} and \ref{fig:r-mssm}\@, we can see one
more important difference between two models. In the ESSM, we can see
that the low-energy bottom-tau ratio is also roughly independent of
the initial value of the top Yukawa coupling for very wide range. This
is because the top Yukawa coupling flows its theoretical infrared
fixed point very quickly almost independently of its initial value,
giving almost the same net effects to the low-energy prediction of the
bottom-tau ratio. Moreover, we should note that we can get the correct
value of the low-energy top quark mass for this all allowed region of
the initial value of the top Yukawa coupling due to its infrared
fixed-point character. These facts are in contrast to the MSSM case,
in which the very large top Yukawa coupling is needed to have proper
value of the bottom-tau ratio. For such a very large top Yukawa
coupling, it flows to its infrared quasi-fixed point
\cite{quasi,b-tau} and leads to small $\tan \beta$ case which makes
the situation much worse than large $\tan \beta$ case. In the ESSM,
the small $\tan \beta$ case is realized for very hierarchical small
bottom Yukawa coupling ($\lsim 10^{-3}$)\@. In this case, the $M_R$
dependence of $R(M_Z)$ is a little increased but it is still rather
smaller than that of the MSSM (Fig.\ \ref{fig:r-essm2})\@. This is
because of the infrared fixed-point behavior of the top Yukawa
coupling and it allows us to have the wide allowed range of the
initial value of $y_t(M_G)$ as well as the large $\tan \beta$
case. However, the small $\tan \beta$ case generally needs a
fine-tuning of Yukawa couplings on the order of $\lsim 10^{-3}$ in the
ESSM.

In conclusion, we have shown that in the ESSM the low-energy
bottom-tau mass ratio is little affected by the presence of the
right-handed neutrinos at the intermediate scale in contrast to the
MSSM case. This result is the consequences of the strong unified gauge
coupling; the infrared fixed-point structure of Yukawa couplings and
the boundary condition of the bottom and tau Yukawa couplings at the
GUT scale. So, it is evident that this is a common feature of strong
coupling unification models. It is interesting that the low-energy
physical parameters can be determined almost independently of the
unknown high-energy physics at the intermediate scale as well as the
GUT scale, and that the allowed parameter region at high-energy scale
can be very wide.

\subsubsection*{Acknowledgements}

One of us (M.\ B.) would like to thank the Aspen Center for Physics
for its hospitality during the middle stage of this work. M.\ B. is
supported in part by the Grant-in-Aid for Scientific Research from
Ministry of Education, Science and Culture and K.\ Y. by the
Grant-in-Aid for JSPS Research fellow.

\newpage

\newpage

\newpage

\end{document}